\documentstyle[aps,amstex,amssymb,epsfig]{revtex}  

\def\eq#1{Eq.~(\ref{#1})}
\def\fig#1{Fig.~\ref{#1}}

\def\tab#1{Tab.~\ref{#1}}

\begin{document}

\title{Stability of bicontinuous cubic phases in 
ternary amphiphilic systems with spontaneous curvature}
\author{U. S. Schwarz${}^{1,2}$ and G. Gompper${}^{1,3}$ \\
${}^{1}$Max-Planck-Institut f\"ur Kolloid- und Grenzfl\"achenforschung, \\
Am M\"uhlenberg, Haus 2, 14476 Golm, Germany\\
${}^{2}$Department of Materials and Interfaces, \\
The Weizmann Institute of Science, Rehovot 76100, Israel \\
${}^{3}$Institut f\"ur Festk\"orperforschung, \\
Forschungszentrum J\"ulich, 52425 J\"ulich, Germany
} 
\date{November 5 1999}

\maketitle

\begin{abstract} 
  We study the phase behavior of ternary amphiphilic systems in the
  framework of a curvature model with non-vanishing spontaneous
  curvature.  The amphiphilic monolayers can arrange in different ways
  to form micellar, hexagonal, lamellar and various bicontinuous cubic
  phases. For the latter case we consider both single structures (one
  monolayer) and double structures (two monolayers).  Their interfaces
  are modeled by the triply periodic surfaces of constant mean
  curvature of the families G, D, P, C(P), I-WP and F-RD. The
  stability of the different bicontinuous cubic phases can be
  explained by the way in which their universal geometrical properties
  conspire with the concentration constraints.  For vanishing
  saddle-splay modulus $\bar \kappa$, almost every phase considered
  has some region of stability in the Gibbs triangle.  Although
  bicontinuous cubic phases are suppressed by sufficiently negative
  values of the saddle-splay modulus $\bar \kappa$, we find that they
  can exist for considerably lower values than obtained previously.
  The most stable bicontinuous cubic phases with decreasing $\bar
  \kappa < 0$ are the single and double gyroid structures since they
  combine favorable topological properties with extreme volume
  fractions.
\end{abstract}


\newpage

\section{Introduction}
\label{sec:intro}

As a function of concentrations and temperature, amphiphilic systems
generically form many different phases, each of which corresponds to a
specific geometrical arrangement of the amphiphilic interfaces
\cite{a:gelb94,a:lipo95c}.  Most ternary amphiphilic systems feature
the disordered micellar, the hexagonal and the lamellar phase. Here
the amphiphilic monolayers, which separate regions of water from
regions of oil, form spheres, cylinders and lamellae, respectively.
However, many of these systems also have stable cubic phases
\cite{a:font90,a:luzz93,a:sedd95}. Near the disordered micellar phase
one often finds a micellar cubic phase and near the lamellar phase a
bicontinuous cubic phase. In the micellar cubic phase, the interfaces
form spheres as they do in the disordered micellar phase, only that
now they are packed in an orderly fashion.  In the bicontinuous cubic
phase, they form sheets which span the whole sample in all directions
of space \cite{a:auss90,a:ptrs96}. Each of these triply periodic
surfaces has a cubic Bravais lattice and divides space into two
unconnected but intertwined labyrinths filled with water and oil,
respectively.

In this work, the polymorphism of ternary amphiphilic systems is
studied in the framework of a curvature model with non-vanishing
spontaneous curvature.  We assume that the shape of the amphiphilic
monolayers is determined by their bending rigidity even in the
presence of concentration constraints.  Therefore we model them by
\emph{surfaces of constant mean curvature}.  This encompasses the
spheres, cylinders and lamellae of the non-cubic phases; for the
bicontinuous cubic phases, the surfaces of constant mean curvature are
triply periodic.  Safran and coworkers considered the same geometries
and showed that the interplay between the bending energy and the
volume constraints can explain some aspects of the generic phase
behavior of ternary amphiphilic systems
\cite{a:safr83,a:safr84,a:wang90}.  In particular, Wang and Safran
\cite{a:wang90} considered the single structure (one monolayer) and
the double structure (two monolayers) from the \emph{D-family}. Their
calculation relied on data for triply periodic surfaces of constant
mean curvature, which were found numerically by Anderson, who
investigated the geometrical properties of the D, P, C(P), I-WP and
F-RD families \cite{a:ande86,a:ande90}.  However, the bicontinuous
cubic phases most often identified experimentally are the various
gyroid structures which correspond to the \emph{G-family}.  Data for
the G-family were calculated only recently by Gro{\ss}e-Brauckmann
\cite{a:gross97,a:gross97b}.  In this work, we consider as
bicontinuous cubic phases single and double structures for \emph{all}
families for which the required geometrical data are known, including
the G-family. Indeed, we find that for negative values of the
saddle-splay modulus $\bar \kappa$, of all bicontinuous cubic phases
considered  
it is mainly the gyroid structures which are stable. They can exist
for values of $\bar \kappa$ which are twice as negative as the values
for which the structures from the D-family have been found to be
stable previously \cite{a:wang90}. We will show that this finding can
be explained nicely by considering the interplay between certain
universal geometrical properties of the various bicontinuous cubic
phases and the volume constraints within the Gibbs triangle.  In fact
we find that the gyroid structures are so favorable because they
have exceptional topological properties and at the same time can
accommodate extreme volume fractions of oil and water.

\section{Curvature model and non-cubic phases}
\label{sec:noncubicphases}

The elastic properties of amphiphilic interfaces are described by the
Canham-Helfrich expression for the elastic energy per unit area
\cite{a:canh70,a:helf73}
\begin{equation} \label{HelfrichHamiltonian} 
f_{elastic} = 2 \kappa\ (H - c_0)^2 + \bar \kappa\ K\ ,
\end{equation}
where $H = (c_1 + c_2) / 2$ and $K = c_1 c_2$ are mean and Gaussian
curvatures, respectively, and $c_1$ and $c_2$ the two principal
curvatures of the surface. The two elastic moduli are the bending
rigidity $\kappa$ and the saddle-splay modulus $\bar \kappa$.  The
main temperature dependence is carried by the spontaneous curvature
$c_0$.  For systems with water, oil and non-ionic surfactants
$C_iE_j$, it is found experimentally that $c_0 \propto (T_b - T)$
where $T_b$ is the balanced temperature \cite{a:stre94}.  Here
positive curvature is defined to be curvature towards the oil regions,
thus oil is the interior phase below the balanced temperature
and water above.  For example, $c_0 \approx 1/(6 l)$ and $c_0 \approx
1/(12 l)$ for $H_2O/C_{14}/C_{12}E_5$ at $T = 20^{\circ} C$ and $T =
34^{\circ} C$, respectively, where $l \approx 1.5\ nm$ is the
amphiphile length in the monolayer and $T_b = 48^{\circ} C$
\cite{a:stre94}. In this work, we assume a positive value for the 
spontaneous curvature, as it is typical for surfactant systems
below the balanced temperature, so that oil
is the interior phase. The tendency to bend towards the oil regions
decreases with temperature since the headgroups' hydration decreases.
For the case of a negative spontaneous curvature, water and oil have
to be interchanged in the various structures as well as in the phase
diagrams presented below.  This is usually the case not only for
surfactants above the balanced temperature, but also for lipid systems
where the monolayers tend to bend towards the water regions due to
their bulky tail regions.  

We consider the dimensionless free energy per unit volume,
\begin{equation} \label{FreeEnergyDensity}
f = \frac{1}{2 \kappa c_0^3 V} \int dA\ f_{elastic} 
= \frac{A}{c_0 V} \left( \frac{H}{c_0} - 1 \right)^2 
- \frac{2 \pi \chi r}{{c_0}^3 V}
\end{equation}
where the integration extends over the neutral surface $A$ of the
amphiphilic monolayers, $V$ is the overall volume, and $r = - \bar
\kappa / 2 \kappa$.  For the first (bending) term in
Eq.~(\ref{HelfrichHamiltonian}), the area integration can be carried
out since we only consider surfaces of constant mean curvature.  For
the second (topological) term, we employ the Gauss-Bonnet theorem,
$\int dA K = 2 \pi \chi$, where $\chi$ is the Euler characteristic of
the surface.  The curvature model is only stable for $-2 \kappa \le
\bar \kappa \le 0$ or $0 \le r \le 1$ \cite{a:helf81}.  Experimentally
there is no straightforward way to measure $\bar \kappa$, but usually
a small negative value is assumed.

The phase diagram at constant temperature is a function of the volume
fractions $\rho_W$, $\rho_O$ and $\rho_A$ of water, oil and
amphiphile, respectively, which are restricted to the Gibbs triangle
by $\rho_W + \rho_O + \rho_A = 1$.  We consider amphiphiles of length
$l$, which have a tail length $\alpha l$ and a head size $(1-\alpha)
l$, with $0 < \alpha < 1$.  Throughout this paper we will use $\alpha
= 1/2$, i.e.\ we will consider amphiphiles which have head and
tail regions of similar size, as they are for $C_{12}E_5$.  The
hydrocarbon volume fraction is given by $v = \rho_O + \alpha \rho_A$
\cite{a:safr84}.  In order to parametrize concentration space, it is
convenient to use $v$ and the ratio $w = \rho_A /
[(\rho_O+\alpha\rho_A) c_0 l]$.  The amphiphile volume fraction
$\rho_A$ is taken to be $A l/ V$.  If the amphiphile is assumed to
occupy the space between the two parallel surfaces at distances
$\alpha l$ and $(1 - \alpha) l$ from the neutral surface, this is a
very accurate approximation when the amphiphile length $l$ is small
compared to the extension of the amphiphilic aggregates, and gives
reasonable results even for the extreme case of oil-free spherical
micelles for $\alpha \simeq 1/2$.

In the following all lengths are measured in units of $l$.
The free-energy densities of the non-cubic phases follow from
\eq{FreeEnergyDensity} as 
\begin{align} \label{FreeEnergyDensityNonCubic_L}
f_L(w,v) & = w v, \\
\label{FreeEnergyDensityNonCubic_C}
f_C(w,v) & = w v \left(\frac{w}{4} - 1\right)^2, \\ 
\label{FreeEnergyDensityNonCubic_S}
f_S(w,v,r) & = w v \left[\left(\frac{w}{3} - 1\right)^2 
                                    - \frac{r w^2}{9}\right] \ .
\end{align}
Note that only $f_S$ depends on $r$, since the other two structures
have no Gaussian curvature.  Since we only consider the free-energy
contribution due to the curvature elasticity of the amphiphilic
monolayers, for the non-cubic phases $f$ has a trivial $v$-dependence.
However, due to close packing, spheres and cylinders have maximal
volume fractions $v_{max} = \sqrt{2} \pi / 6 = 0.74$ and $\pi / 2
\sqrt{3} = 0.91$, respectively.  With $w$, $v$ and $r$, the model has
a three-dimensional parameter space.

It can be seen from \eq{FreeEnergyDensityNonCubic_L} that the Maxwell
construction is not possible for this model, since $f_L$ is not
concave.  Therefore we use the intersections of the free-energy
densities of different phases as an indication for the location of
phase transitions.  We want to remark parenthetically that Wang and
Safran \cite{a:wang90} considered the free energy per unit area
(rather than per unit volume); this amounts to an overall factor of $w
v / c_0$ (the dimensionless specific area) in the free-energy density
of all phases.  As long as the location of the phase transitions is
estimated from the intersection of the free-energy curves, the two
approaches are equivalent. However, for a calculation of two-phase
regions, the use of the free energy per unit \emph{volume} becomes
essential --- since in general coexisting phases will have different
amphiphile concentrations.

For $r = 0$ (vanishing saddle-splay modulus) and $v < 0.74$ (no
excluded volume effects), Eqs.~(\ref{FreeEnergyDensityNonCubic_L}),
(\ref{FreeEnergyDensityNonCubic_C}) and
(\ref{FreeEnergyDensityNonCubic_S}) imply the phase sequence $L \to C
\to S$ with decreasing $w$.  $f_S$ begins to rise again for $w \le 3$.
This is identified with the \emph{emulsification failure}, the
coexistence of $S$ and an excess oil phase at low amphiphile
concentration \cite{a:wang90}. In this paper, we define the
emulsification failure not as the minimum in $f_S$ in regard to $w$,
but by using a Maxwell construction in regard to $w$ between $S$ and
an excess oil phase, i.e.\ by solving $\partial f_S/\partial w = f_S /
w$ for $w$.  With increasing $r$ (and $v < 0.74$), spheres become more
favorable, while cylinders are increasingly suppressed and finally
disappear for $r > 1/4$.  In particular, the phase boundaries $S-C$,
$S-L$, $C-L$ and the emulsification failure are obtained from
Eqs.~(\ref{FreeEnergyDensityNonCubic_L}),
(\ref{FreeEnergyDensityNonCubic_C}) and
(\ref{FreeEnergyDensityNonCubic_S}) to be $w = 24 / (7 - 16 r)$, $w =
6 / (1 - r)$, $w = 8$ and $w = 3/(r - 1)$, respectively.

It is important to realize that not all values of $(v,w)$ are
physically relevant.  In \fig{Mapping} we show the mapping between the
$(v,w)$-plane and the Gibbs triangle for ternary mixtures.  The binary
limits O-W, W-A and A-O of the Gibbs triangle correspond to $\rho_O +
\rho_W = 1$, $\rho_W + \rho_A = 1$ and $\rho_A + \rho_O = 1$,
respectively.  In the $(v,w)$-plane, this corresponds to the lines $w
= 0$, $w=1/(\alpha c_0)$ and $w = (1/v - 1) /((1-\alpha) c_0)$,
respectively. The line $v = 0$ is mapped onto the W-apex.
\fig{Mapping} demonstrates (for $\alpha = 1/2$) this mapping of the
$(v,w)$-plane onto the Gibbs triangle for $c_0 = 1/6$ and $c_0 =
1/12$.  As mentioned above, these values correspond to the system
$H_2O/C_{14}/C_{12}E_5$ at $T = 20^{\circ} C$ and $T = 34^{\circ} C$,
respectively \cite{a:stre94}.  Smaller/larger values for the
spontaneous curvature (which corresponds to higher/lower temperatures
or other components) increases/decreases the size of the relevant
region in \fig{Mapping}a to larger/smaller values of $w$. One also can
see in \fig{Mapping}b that for $\alpha = 1/2$ lines of constant $v$
are perpendicular to the W-O side and lines of constant w are straight
lines through the W-apex.

\section{Bicontinuous cubic phases}
\label{sec:cubicphases}

\subsection{Properties of constant-mean-curvature surfaces}

We model the amphiphilic monolayers in the bicontinuous cubic phases
by triply periodic surfaces of constant mean curvature $H = (c_1 + c_2
)/2$.  For the special case $H = 0$, this leads to triply periodic
minimal surfaces with $c_1 = - c_2$ and $K = c_1 c_2 \le 0$.  Due to
the Gauss-Bonnet theorem, $\int dA K = 2 \pi \chi$, these surfaces
have a negative Euler characteristic $\chi$ per unit cell.  Triply
periodic minimal surfaces are commonly used to model all kinds of
extended sheet-like structures in condensed matter systems, in
particular the mid-surfaces of the lipid bilayers in \emph{inverse
cubic phases} which are very prominent in lipid-water mixtures
\cite{a:auss90,a:ptrs96}.  Before 1970 only three cubic triply
periodic minimal surfaces have been known (D, P and C(P))
\cite{a:nits89,a:dier92}.  Then Schoen described five more (G, F-RD,
I-WP, O,C-TO and C(D)) \cite{a:scho70}.  Today some more examples are
known \cite{a:karc96,a:fisc96}, but none of them seems to be of
physical relevance.  Karcher proved in 1989 not only the existence of
the triply periodic minimal surfaces described by Schoen, but also
that the simpler of them can be deformed into triply periodic surfaces
of constant (non-zero) mean curvature \cite{a:karc89}.  As for any
triply periodic surface, space is divided into two percolating
labyrinths.  A shift of $H$ to positive/negative values shrinks/expands
one labyrinth, while it expands/shrinks the other.  Thus two
branches are generated, which both end in cubic arrangements of
(infinitesimally connected and possibly self-intersecting) spheres.
Since the Euler characteristic is a topological quantity connected to
the genus $g$ of the surface by $\chi = 2 (1 - g)$, it does not change
within a family.  Anderson studied the cubic families D, P, C(P), I-WP
and F-RD and calculated for a conventional unit cell of unit lattice
constant both the volume fraction $v$ of one labyrinth as a function
of scaled mean curvature $H^*$ (the volume fraction of the other
labyrinth follows as $1 - v$) and the scaled surface area $A^*$ as a
function of $v$ \cite{a:ande86,a:ande90}.  Recently $v(H^*)$ and
$A^*(v)$ were calculated also for the G-family by Gro{\ss}e-Brauckmann
\cite{a:gross97,a:gross97b}.

In this work we consider the families G, D, P, C(P), I-WP and F-RD,
for which these geometrical data are available. Their minimal surface
member is shown in \fig{PicturesSingle} for D, C(P), I-WP and F-RD
in one conventional unit cell. The numerically calculated data points
for $v(H^*)$ and $A^*(v)$ per conventional unit cell are taken from
Refs.~\cite{a:ande86,a:gross97}; rearrangement and interpolation with
cubic splines provides smooth functions $v(H^*)$, $H^*(v)$ and
$A^*(v)$.  The data are used only up to the extremal values of $v$,
beyond which the curves bend backwards. Beyond these points, the
surfaces resemble ensembles of spheroidal regions connected by nearly
unduloidal necks, which we do not consider to be of physical relevance
\cite{a:ande90b}.  The functions $H^*(v)$ and $A^*(v)$ as used in this
work are plotted in \fig{TPHS} for the six families considered.  In
fact, only one of the two curves for each family carries independent
information since the other one can be constructed by using $dA^* = 2
H^* dv$ \cite{a:ande90}.  Each family only exists over a certain range
of volume fractions; the extreme cases are the G-family with $v\in
[0.056,0.944]$ and the C(P)-family with $v\in [0.481,0.519]$.  If the
minimal surface member of a family divides space into two congruent
labyrinths, they have the same volume fraction $v_0 = 1/2$, the two
branches are symmetric to each other and the minimal surface as well
as the family itself is called \emph{balanced}.  For the structures
considered here, this is the case for G, D, P and C(P).  For these
structures, the curves $H^*(v)$ and $A^*(v)$ are symmetrical with
respect to $v = 1/2$.  For the non-balanced families I-WP and F-RD,
the two labyrinths are of different topology and $v_0 \neq 1/2$.  In
\tab{DataTable}, we collect data connected to the minimal surface
members of each family, which we will use later in our discussion.

In the vicinity of the minimal surface, the volume fraction $v$ as a
function of scaled mean curvature $H^*$ can be approximated as $v(H^*)
= v_0 - c H^* + {\cal O}({H^*}^2)$. With $dA^* = 2 H^* dv$ it follows that
\begin{align} \label{TPHSApprox}
H^*(v) & = - \frac{(v - v_0)}{c}       + {\cal O} \left( (v-v_0)^2 \right) \\
A^*(v) & = A_0 - \frac{(v - v_0)^2}{c} + 
                               {\cal O} \left( (v-v_0)^3 \right) \nonumber 
\end{align}
near the minimal surface.  The values for $A_0$, $v_0$, $c$ are
given in \tab{DataTable}.  Since the surfaces of constant mean
curvature can be expected to have similar geometrical properties as
parallel surfaces in the vicinity of a minimal surface, the magnitude
of $c$ can be estimated as follows. If $t$ denotes the perpendicular
distance from the minimal to its parallel surface, to lowest order in
$t$ the volume fraction $v$ and the mean curvature $H^*$, {\it
averaged} over the surface in the whole unit cell, are given by $v =
v_0 + A_0 t$ and $H^* = 2 \pi \chi t / A_0$, respectively
\cite{a:spiv79,a:hyde89}.  Thus $H^* = 2 \pi \chi (v - v_0) / {A_0}^2$
and $c = - {A_0}^2 / 2 \pi \chi$ for the parallel surface case.  The
corresponding numbers are given in \tab{DataTable}; except for C(P),
the overall agreement with the numerical data for $c$ is remarkably
good.

\subsection{Single and double structures}

The simplest case of a cubic bicontinuous phase in a ternary
amphiphilic system is a \emph{single structure} where the amphiphilic
monolayers form \emph{one} triply periodic surface.  Then one
labyrinth is filled with oil and the other with water.  For balanced
families, filling either of the two labyrinths with oil gives the same
single structure; for the non-balanced families I-WP and F-RD, this
yields different single structures which we denote by I, WP, F and RD,
respectively \cite{a:ande90b}.  Here the symbols I and F correspond to
the curves plotted for I-WP and F-RD in \fig{TPHS}.  Since mean
curvature is defined as curvature towards the oil regions, WP and RD
follow by the replacements $v \to 1 - v$ and $H \to - H$ from the data
of I and F plotted in \fig{TPHS}, respectively.  Thus, altogether we
consider 8 different single structures, which exist for the volume
intervals $[0.056,0.944]$ for G, $[0.131,0.869]$ for D,
$[0.249,0.751]$ for P, $[0.481,0.519]$ for C(P), $[0.357,0.857]$ for
I, $[0.143,0.643]$ for WP, $[0.439,0.625]$ for F and $[0.375,0.561]$
for RD.

To each single structure corresponds a \emph{double structure} where
the amphiphilic monolayers form \emph{two} triply periodic surfaces
arranged roughly parallel and on either side of the minimal surface of
the corresponding single structure.  Both surfaces divide space into
two labyrinths which are topologically equivalent to those of the
initial structure.  However, since two amphiphilic monolayers are
present, now these labyrinths are filled with the same component and
separated by a bilayer which is filled with the other component.
Single and double structures are also known as monolayer and bilayer
structures \cite{a:ande90b,a:hyde89}.  A double structure can be
either of type I (water-filled bilayer) or of type II (oil-filled
bilayer) \cite{a:luzz68}.  There will be no problem below to
distinguish the symbol I for double structure of type I from the
symbol I for one of the two simple structures of the I-WP family.  In
\tab{singledouble}, we summarize the classification of bicontinuous
cubic structures.  In the following, single structures, type I double
structures and type II double structures are abbreviated as Q, Q${}_I$
and Q${}_{II}$, respectively. Q${}_{II}$ structures are also known as
\emph{inverse bicontinuous cubic phases}. In \fig{PicturesDouble} we
show single and double gyroid structures in one conventional unit
cell.  For many surfactant-water and lipid-water systems, G${}_{II}$
and/or D${}_{II}$ are well established; moreover there are reports on
P${}_{II}$ and G${}_I$ structures in these binary amphiphilic systems
\cite{a:font90,a:luzz93,a:sedd95}.  Q structures in binary systems
have been discussed only theoretically so far
\cite{a:gomp92b,a:Gomp95a,a:linh98}.  For many systems with water, oil
and surfactant, stable bicontinuous cubic phases have been reported,
however often without identification of their space group (see e.g.\ 
Ref.~\cite{a:leav95} for the system $H_2O/C_{10}/C_{12}E_5$).  The
best established identification is a number of Q$_{II}$ structures for
the system DDAB-water-styrene \cite{a:stro92}.  There are a few
reports on Q structures \cite{a:ande86,a:raed89}, and none on
Q${}_{I}$ structures.  However, several speculations on single
structures can be found in the literature (e.g.\ in
Refs.~\cite{a:hyde89,a:mari88}), and our recent theoretical work on
ternary systems with vanishing spontaneous curvature suggests that
they should have some physical relevance \cite{a:schw98a}.  To our
knowledge, hardly nothing is known on bicontinuous cubic phases in
ternary systems with water, oil and lipid.

In order to construct a double structure from a given family of triply
periodic surfaces of constant mean curvature, we take two surfaces
corresponding to $H$ and $-H$. Thus there is one Q${}_{I}$ and one
Q${}_{II}$ structure for each of the 6 families.  For Q${}_{I}$
structures, the minimal surface case corresponds to $v = 1$.  As both
surfaces accumulate mean curvature in their respective branches, the
volume fraction decreases until the first labyrinth reaches its
minimal size (which is the minimal volume fraction of the
corresponding single structure). The volume intervals covered by the
Q${}_{I}$ structures turn out to be $[0.112,1.0]$ for G${}_I$,
$[0.262,1.0$] for D${}_I$, $[0.498,1.0]$ for P${}_I$, $[0.962,1.0]$
for C(P)${}_I$, $[0.624,1.0$] for I-WP${}_I$ and $[0.818,1.0$] for
F-RD${}_I$. The Q${}_{II}$ structures follow from the corresponding
Q${}_{I}$ structures by an interchange of oil and water.  Thus here
the minimal surface case corresponds to $v = 0$, and the volume
fractions covered are complementary to the ones given for the
Q${}_{I}$ structures.

\subsection{Curvature and topology index}

For bicontinuous cubic structures, the free-energy density depends on
the hydrocarbon volume fraction $v$ in a non-trivial way, since only
one or two amphiphilic aggregates are present. The lattice constant is
denoted with $a$.  For a given value of $v$, the mean curvature
$H(v,a)=H^*(v)/a$ and the surface area $A(v,a)=A^*(v) a^2$ within a
unit cell are determined by the curves plotted in \fig{TPHS}. The
amphiphile concentration $\rho_A = A(v,a)/a^3 = A^*(v)/a$ fixes $a$,
so that $a = A^*(v) / \rho_A = A^*(v) / (w v c_0)$.  Using the
Gauss-Bonnet theorem for the Gaussian curvature term, we find from
\eq{FreeEnergyDensity} for the free-energy density of a bicontinuous
cubic structure
\begin{align} \label{FreeEnergyDensityCubic}
f_{BC}(w,v,r) & = \frac{A}{c_0 V} \left(\frac{H^*(v)}{c_0\ a} - 1\right)^2 
- \frac{2 \pi \chi r}{(c_0\ a)^3} \nonumber \\
& = w v \left[\Lambda(v)\ w v - 1\right]^2 
+ r\ \frac{(w v)^3}{\Gamma(v)^2}
\end{align}
with 
\begin{equation} \label{GeometricalProperties}
\Lambda(v) = \frac{H^*(v)}{A^*(v)}\ , \quad 
\Gamma(v) = \left(\frac{A^*(v)^3}{2 \pi |\chi|}\right)^{1/2}\ .
\end{equation}
Here $\chi$ denote the Euler characteristic of the interfaces within
one conventional unit cell, as given in \tab{DataTable}.  The
\emph{curvature index} $\Lambda(v)$ and the \emph{topology index}
$\Gamma(v)$ are two universal geometrical quantities which
characterize a surface in three-dimensional space.  Their significance
can be understood rather easily from the observation that both the mean
and Gaussian curvatures can be made dimensionless by multiplying them
with appropriate powers of $V/A$, which is the only relevant length
scale; this implies $H (V/A) = \Lambda$ and $|K| (V/A)^2 = 2 \pi
|\chi| V^2 / A^3 = 1 / \Gamma^2$.  It is not difficult to see that
both quantities are not only invariant under scale transformations,
but also under a change of unit cell. They also occur in integral
geometry, a mathematical theory concerned with invariant geometrical
measures \cite{a:schn93}. For convex bodies in three-dimensional
space there exist two independent isoperimetric inequalities and
therefore two independent isoperimetric ratios, which usually are
chosen to be $\Lambda$ and $1 / \Gamma^2$. The equivalence between
these quantities and the isoperimetric ratios holds since
for surfaces of constant mean curvature, the integral mean curvature,
which is one of the Minkowski functionals of integral geometry, can be
replaced by $H A$. The curvature index $\Lambda$ describes how
strongly the structures is curved and the topology index $\Gamma$
describes its porosity (the larger its value, the less holes the
structure has). For minimal surfaces, $\Lambda$ vanishes and $\Gamma$
remains the only relevant quantity. Its significance for amphiphilic
systems has been pointed out by Hyde \cite{a:hyde89}, and its
variation as a function of crystallographic determinants has been
studied recently by Fodgen and Hyde \cite{a:fodg99}. For
multicontinuous structures build from $n$ sheets, the curvature index
is smaller and the topology index is larger by a factor of $n$
compared to the corresponding single structure.  In particular, for
double structures the curvature index is half and the topology index
is twice as large as for single structures.  Thus, double structures
are less porous than single structures since they have disconnected
surfaces.  We can infer from \tab{DataTable} that the minimal gyroid
is the least porous of the single structures with $\Gamma = 0.7667$.
We want to remark parenthetically that --- surprisingly ---
bicontinuous random surfaces have similar values of $\Gamma$ as single
cubic structures.  For example, it can be shown exactly that the
isosurfaces of Gaussian random fields with $\langle H \rangle = 0$
have $\Gamma = \sqrt{8} / \pi = 0.9003$ \cite{a:teub91,a:Gomp95}.
Since this value is only slightly larger than that of the single
gyroid phase, it can be concluded that the random sponge phase on
average features only few disconnected or multiple sheets.  However,
when the random sponge's interfaces are made to acquire curvature, the
topology index grows strongly with $\Lambda$ since disconnected parts
proliferate \cite{a:Gomp95}.  The opposite is true for the families of
surfaces of constant mean curvature considered here; if the mean
curvature is increased from zero, the topology remains the same, but
the area content decreases.  Therefore the topology index decreases,
since for given topology it measures the specific surface area.

In order to evaluate the free-energy density of the bicontinuous cubic
phases from \eq{FreeEnergyDensityCubic} as a function of $w$, $v$ and
$r$, curvature and topology index as defined in
\eq{GeometricalProperties} have to be calculated as a function of $v$.
For the 8 Q structures, they follow in a straightforward way from the
data plotted in \fig{TPHS}.  In order to derive them for the 6
Q${}_{I}$ structures, we first consider the non-balanced case, thus
the two labyrinths $1$ and $2$ have different topologies (e.g.\ for
I-WP, we have $1 = I$ and $2 = WP$).  We first construct $v_{I}(H^*) =
v_1(H^*) + v_2(H^*)$ and then invert it to obtain $H^*_I(v)$, the mean
curvature of both surfaces as a function of the overall volume
fraction (which is distributed onto both labyrinths).  The surface
area then follows as $A^*_I(v) = A^*_1(v_1(H^*_I(v))) +
A^*_2(v_2(H^*_I(v)))$.  The Euler characteristic is $\chi_I = \chi_1 +
\chi_2$ (with $\chi_1 = \chi_2$ given in \tab{DataTable}).  With
$H^*_I(v)$, $A^*_I(v)$ and $\chi_I$ calculated, we then can evaluate
$\Lambda$ and $\Gamma$ in \eq{GeometricalProperties} for the
non-balanced Q${}_{I}$ structures.  For the non-balanced Q${}_{II}$
structures, we have to evaluate them using $H^*_{II}(v) = -
H^*_I(1-v)$, $A^*_{II}(v) = A^*_I(1-v)$ and $\chi_{II} = \chi_I$.  For
the balanced families, one has $v_1 = v_2 = v/2$, $H^*_I(v) =
H^*(v/2)$, $A^*_I(v) = 2 A^*(v/2)$, $H^*_{II}(v) = - H^*((1-v)/2)$ and
$A^*_{II}(v) = 2 A^*((1-v)/2)$.  This amounts to using $\Lambda_I(v) =
\Lambda(v/2)/2$, $\Gamma_I(v) = 2 \Gamma(v/2)$, $\Lambda_{II}(v) = -
\Lambda((1-v)/2)/2$ and $\Gamma_{II}(v) = 2 \Gamma((1-v)/2)$ in
\eq{GeometricalProperties}. Note the minus sign for the curvature of
the Q${}_{II}$ structures since the monolayers change their
orientation compared to the Q${}_{I}$ structures.  In
\fig{IndexesPlot} we plot $\Lambda$ and $\Gamma$ as a function of $v$
for all single and double structures considered.  Recall that the
minimal surface case corresponds to $v = v_0$, $1$ and $0$ for Q,
Q${}_{I}$ and Q${}_{II}$ structures, respectively; in these cases, the
curvature index disappears and the topology index acquires its maximum
value.  In \tab{DataTable}, we give the values of $\Gamma_0 = \Gamma(v
= v_0)$ for all families considered.  This implies the hierarchy G, D,
I-WP, P, C(P), F-RD with decreasing $\Gamma_0$.  From
\fig{IndexesPlot} we also see that the single structures I and WP in
fact can become better than single D for certain values of $v$.

We see from \fig{IndexesPlot} that single structures for $v > v_0$ and
all Q${}_{II}$ structures have negative values for the curvature index
$\Lambda$. From \eq{FreeEnergyDensityCubic} it therefore follows that
these structures have $f_{BC} > w v = f_L$, compare
\eq{FreeEnergyDensityNonCubic_L}; thus they are always less stable
than the lamellar phase. In other words, since we consider the case
that the monolayers prefer to bend towards the oil regions (positive
spontaneous curvature $c_0$), in the framework of the curvature model
no phase can be stable which curves towards the water regions.  One
should note, however, that single structures for $v > v_0$ or
Q${}_{II}$ structures could occur even for positive $c_0$ if they were
stabilized by other contributions to the free energy which are
neglected in our treatment. For the case of negative $c_0$, the
situation is reversed: now the monolayers prefer to bend towards the
water regions, and from the double structures the (inverse) Q${}_{II}$
and not the Q${}_{I}$ structures are stable.  In general, for ternary
systems one expects Q in the middle of the Gibbs triangle (possibly in
the vicinity of a microemulsion phase for surfactant systems), Q$_{I}$
near the binary side amphiphile-oil and Q${}_{II}$ near the binary
side amphiphile-water. For surfactant systems, which usually have $c_0
> 0$ at room temperature, the curvature model predicts the stability
of Q and Q$_{I}$ structures; for lipid systems, which usually have
$c_0 < 0$ at room temperature, Q and (inverse) Q${}_{II}$ structures
are predicted.  This general prediction conforms with the predominance
of inverse phases for lipid-water mixtures. Note that the surfactant
systems DDAB-water-styrene \cite{a:stro92} mentioned above is an
exception to our distinction between surfactant and lipid systems
since the two tails result in $c_0 < 0$ and therefore lead to inverse
phases like in the lipid case.

\section{Phase behavior}
\label{sec:phasebehavior}

\subsection{Phase diagrams for $r = 0$}

Altogether we consider 17 different phases, 3 non-cubic, 8 single and
6 double structures of type I.  Moreover there always exists an
emulsification failure at low amphiphile-to-oil ratios.  We first
discuss the case of vanishing saddle-splay modulus, $r = 0$.  For the
bicontinuous cubic phases we see from \eq{FreeEnergyDensityCubic} that
the optimal value $f = 0$ for the free-energy density is achieved for
$w(v) = 1 / [v \Lambda(v)]$.  For the spherical and cylindrical phases
we find $w(v) = 3$ and $w(v) = 4$ from
Eqs.~(\ref{FreeEnergyDensityNonCubic_S}) and
(\ref{FreeEnergyDensityNonCubic_C}).  For each phase the corresponding
line $w(v)$ lies in the middle of its region of stability; these lines
are plotted in \fig{PhaseDiagramRfixed}a instead of the full phase
diagram for $r = 0$.  In the following we will denote them as
\emph{lines of vanishing frustration} since they mark the specific
parameter values for which a given phase can satisfy both bending and
concentration constraints simultaneously. In the $(v,w)$-plane, each
of the structures considered has such a line, thus each of them has
some region of stability where its particular geometry serves best to
accommodate the volume fractions of the different components.  In fact
there are even several values of $(w,v)$ where the free-energy density
of different structures is degenerate.  In \fig{PhaseDiagramRfixed}b
we map the lines of vanishing frustration from
\fig{PhaseDiagramRfixed}a onto the Gibbs triangle for $c_0 = 1/6$. For
these values, the Q structures run towards the W-A side and the
Q${}_{I}$ structures towards the A-O side.  Although the lines of
vanishing frustration for C(P), C(P)${}_{I}$ and F-RD${}_{I}$ are not
mapped onto the Gibbs triangle, the phase behavior here remains highly
degenerate.

\fig{PhaseDiagramRfixed}a demonstrates that the different structural
types considered occupy different regions of the phase diagram in a
very characteristic fashion: for large $w$, Q and Q${}_{I}$ structures
are stable for $v \lesssim 1/2$ and $v \lesssim 1$, respectively.
With decreasing $w$, the regions of stability curve to the left.  In
order to understand the sequence of phases within the band-like region
of each structural type for large $w$, it is useful to expand the
free-energy densities of the various bicontinuous structures about the
minimal surface members by using \eq{TPHSApprox}.  For the Q and
Q${}_{I}$ structures, this corresponds to an expansion about the
volume fractions $v_0$ and $1$, respectively.  Again we only consider
the lines of vanishing frustration for which $f = 0$ and $w(v) = 1 /
[v \Lambda(v)]$. The curvature index $\Lambda(v)$ defined in
\eq{GeometricalProperties} can be approximated by
\begin{align} \label{GeometricalPropertiesApprox} 
\Lambda(v)      & = \frac{- (v -  v_0)}{c A_0} 
                 + {\cal O} \left( (v-v_0)^2 \right)\ , \nonumber \\
\Lambda_I(v)    & = \frac{(1 - v)}{4 c A_0} 
                 + {\cal O} \left( (1-v)^2 \right)\ .  
\end{align}
In the same order of approximation, the lines of vanishing frustration
then follow as $w = - c A_0 / v_0 (v - v_0)$ and $w = - 4 c A_0 / (v -
1)$ for Q and Q${}_{I}$ structures, respectively.  Thus the minimal
surface case corresponds to the stable solutions for $w \gg 1$ at $v =
v_0$ and $v = 1$, respectively, where the curvature indexes of the
corresponding structures disappear (compare \fig{IndexesPlot}).  For
the balanced single structures and the double structures the hierarchy of
the different phases within the band-like region occupied by a certain
structural type is thus determined by the values of $c A_0$, which are
given in \tab{DataTable} for the 6 families considered.  The only
exception are the 4 non-balanced single structures, which cannot be
compared in this way, since the value for $v_0$ is different for each
of them.  Using the approximation $c \approx - {A_0}^2 / 2 \pi \chi$
derived above, we find $c A_0 \approx {\Gamma_0}^2$, so that the
sequence is approximately determined by the topology index of the
minimal-surface member of each family.  In fact the structures in
\tab{DataTable} are ordered with decreasing $\Gamma_0$.  In
particular, for a given structural type and values of the hydrocarbon
volume $v$ outside the (rather restricted) $v$-intervals of existence
of I-WP, F-RD and C(P), we expect from
Eq.~(\ref{GeometricalPropertiesApprox}) to find the sequence G - D - P
as a function of either $v$ or $w$.

\subsection{Phase diagrams for $r > 0$}

When the saddle-splay modulus becomes negative, so that $r>0$, the
free-energy densities of lamellae and cylinders do not change since
these structures have vanishing Gaussian curvature.  However, spheres
have positive Gaussian curvature and therefore their free-energy
density decreases (compare \eq{FreeEnergyDensityNonCubic_S}).  Since
bicontinuous cubic phases have negative (integral) Gaussian curvature,
their free-energy density increases (compare
\eq{FreeEnergyDensityCubic}) and they will be suppressed towards large
$w$ by the lamellar phase and towards small $w$ by the cylindrical
phase; these phases in turn will for sufficiently large $r$ be
suppressed by the spherical phase. Since ternary amphiphilic systems
presumably have negative but small values of the saddle-splay modulus,
the important questions here are up to which value of $r$ the
bicontinuous cubic phases remain stable, and which of the 14 different
structures considered performs best.  It follows from
\eq{FreeEnergyDensityCubic} that the relevant quantity is the topology
index: the larger its value for a certain bicontinuous cubic structure
which is stable for $r = 0$, the longer this structure stays stable
with increasing $r$.  \fig{IndexesPlot} shows that for any value of
$v$, the double structures have larger geometry indices than single
structures.  Within each of the two relevant structural classes, it is
the gyroid structure which has the largest value of the topology index
(compare \tab{DataTable}). We therefore conclude that the double
gyroid G${}_{I}$ should dominate phase behavior for $r > 0$ for
topological reasons.  However, there are three restrictions to this
general conclusion.  First, G${}_I$ can only realize $v \in
[0.112,1.0]$.  Second, before G${}_{I}$ can dominate all other
bicontinuous cubic phases with increasing $r$, it might be already
dominated itself by the lamellar phase L and the cylindrical phase C.
And third, since the topological term is weighted by a factor $(w
v)^3$ in \eq{FreeEnergyDensityCubic}, the double gyroid cannot perform
so well for small $v$ as it can for larger $v$.

Our numerical results nicely corroborate this analysis and shows the
exact outcome of the balance between the different principles
mentioned.  In \fig{PhaseDiagramVfixed} we show phase diagrams as a
function of $w$ and $r$ for $v = 0.1$, $0.2$, $0.4$ and $0.6$. Since
$w$, $v$ and $r$ define the parameter space, these figures show nearly
the complete phase behavior predicted by the model. Only the
subsequent mapping onto the Gibbs triangle is affected by the chosen
values for $\alpha$ and $c_0$. In \fig{PhaseDiagramVfixed}, we draw
the lines of constant $w$, which marks the upper edge of the part of
the $(r,w)$-plane which is mapped onto the Gibbs triangle for $\alpha
= 1/2$ and $c_0=1/6$.  The degeneracy of the bicontinuous cubic phases
for $r = 0$ discussed above disappears quickly with increasing $r$.
They eventually all disappear because the lamellar phase L and the
cylindrical phase C become more stable. At $r = 0.25$, C itself is
suppressed by the spherical phase S. The double gyroid G${}_I$ remains
stable for larger values of $r$ than all other bicontinuous cubic
phases. In fact for $v = 0.2$ (\fig{PhaseDiagramVfixed}b) it is stable
up to $r = 0.2$. This result stands in marked contrast to the results
of Ref.~\cite{a:wang90}, which predicted that the most stable phase
should be D${}_{I}$ with a stability limit of $r = 0.1$. For $v = 0.1$
(\fig{PhaseDiagramVfixed}a) the double gyroid cannot exist and the
single gyroid G is the only stable bicontinuous cubic phase.  For $v =
0.6$ (\fig{PhaseDiagramVfixed}d) the single structures are not stable
since their curvature index is negative for $v > v_0$ (compare
\fig{PhaseDiagramRfixed}a). If more than one structure is stable
within one structural class, we see the sequence G - D - P which was
shown above to be determined by the topology index as well.

In \fig{PhaseDiagramGibbs} we show the Gibbs triangles for $r = 0.01$
and $c_0 = 1/6$, $r = 1/15 = 0.07$ and $c_0 = 1/6$, as well as $r =
1/15$ and $c_0 = 1/12$.  In the first case of a very small value of
the saddle-splay modulus, one still sees the degeneracy of the case $r
= 0$. For the more negative value, only the single gyroid G and the
double gyroid G${}_I$ are stable. Comparing with
\fig{PhaseDiagramRfixed}, we see that the G-phase is stable near its
extremal volume fraction of $v=0.056$, while the G$_I$-phase is stable
for a large range of volume fractions, which covers the region of the
Gibbs triangle where its line of vanishing frustration is located for
$r = 0$. The effect of decreasing spontaneous curvature from $c_0 =
1/6$ in \fig{PhaseDiagramGibbs}b to $c_0 = 1/12$ in
\fig{PhaseDiagramGibbs}c is to extend the region of stability of the
lamellar and bicontinuous phases away from the W-A side, towards the
center of the Gibbs triangle.

\section{Discussion and Conclusions}
\label{sec:conclusion}

In order to calculate phase behavior of ternary amphiphilic systems,
we investigated a simple curvature model with non-zero spontaneous
curvature. In particular, we focused on the bicontinuous cubic phases
whose interfaces were modeled by triply periodic surfaces of constant
mean curvature. We showed that for this class of surfaces, the
free-energy density of the bicontinuous cubic phases can be written in
a very general form which emphasizes the universal geometrical
character of the problem studied.  It consists of two terms: the
bending term depends on the curvature index $\Lambda$ and the
topological term depends on the topology index $\Gamma$. The relative
strength of the two terms is determined by $r = - \bar \kappa / 2
\kappa$.  The relevance of the two quantities $\Lambda$ and $\Gamma$
not only depends on $r$, but also on the way by which they are
weighted in the free-energy density by the two variables $w$ and $v$
which parametrize concentration space.  Several properties have to
conspire for a specific phase to be stable at a certain point of the
phase diagram. First, its geometrical properties have to allow to
accommodate the given concentrations (this imposes some constraints on
the allowed values for the hydrocarbon volume fraction $v$). Second, the
resulting mean curvature must be close to the given spontaneous
curvature in order to keep the bending term small.  Since there are
two independent degrees of freedom in concentration space, one of
which is sufficient to adjust the mean curvature to its optimal value,
the regions of stability are at least one-dimensional in the
$(v,w)$-plane for $r=0$. The exact location of these lines of
vanishing frustration is determined by the curvature index
$\Lambda(v)$. Phases with negative curvature index $\Lambda$ are less
stable than the lamellar phase. For positive spontaneous curvature
(monolayer bending towards oil regions), this rules out the single
structures for $v > v_0$ and the double structures of type II.
Third, for $r > 0$ the structure has to be favored by the topological
term also. This requires large values for the topology index
$\Gamma(v)$ and basically favors the double gyroid G${}_I$.  Fourth,
the concentrations in the regions of stability have to be physically
relevant, i.e.\ they have to correspond to those of the Gibbs
triangle. The mapping of the phase diagram as a function of $v$ and
$w$ onto the Gibbs triangle depends on specific values for the
amphiphile chain length $\alpha$ and the spontaneous curvature $c_0$
and cuts off some of the phase behavior in the $(v,w)$-plane.

The phase behavior for $r=0$ is highly degenerate.  This degeneracy
has been discussed already in Ref.~\cite{a:Brui92} for the
Canham-Helfrich Hamiltonian without spontaneous curvature and
concentration constraints.  In order to resolve the question of the
relative stability of the different bicontinuous cubic phases, one has
to consider further physical effects like topological contributions
($r > 0$), van der Waals, electrostatic or steric interactions, higher
order curvature terms \cite{a:Brui92} or packing energies for the
hydrocarbon chains \cite{a:ande88,a:dues97}.  Our analysis shows that
although every phase considered has some region of stability in phase
space, there are certain general principles which allow to understand
the complicated structure of the resulting phase diagram.  The
locations of the regions of stability are determined by the curvature
index $\Lambda$; we have derived a simple approximation for $\Lambda$,
which is valid for large $w$ and explains why Q and Q${}^{I}$
structures are stable for $v \lesssim 1/2$ and $v \lesssim 1$, 
respectively.  Moreover, it turns out that the
sequence of phases within the band-like regions of a certain
structural type is determined by the value of $c A_0$, which in turn
can be well approximated by ${\Gamma_0}^2$.  Thus the relative
location of the different phases of one type is determined by the
topology index of the minimal-surface member of that family.  

When the saddle-splay modulus becomes negative, more and more of the
cubic bicontinuous phases disappear until eventually all of them are
suppressed by the non-cubic phases. However, we found that the
bicontinuous cubic phases remain stable for considerably higher values
of $r$ than found previously \cite{a:wang90}.  Since the structure
performs best which has both a high topology index and can accommodate
large ranges of hydrocarbon volume fraction $v$, the double gyroid
G${}_I$ becomes the most stable bicontinuous cubic phase for
increasing $r$. Note that G${}_I$ is the only double structure of type I   
where the oil-filled labyrinths consist of channels, which are connected  
by junctions of three-fold coordination exclusively. Its outstanding 
stability with respect to $r$ can therefore be explained 
by the fact that it is the most cylinder-like of
the bicontinuous cubic phases. The same geometrical property in fact
can stabilize also the entropy-dominated microemulsion since
low-coordinated vertices provide a lot of configurational entropy
\cite{a:tlus97}. Since with increasing $r$ the double gyroid G${}_I$ 
is eventually suppressed by the lamellar and the cylindrical
phase before it can suppress all other bicontinuous cubic phases,
the single gyroid G (with has the most favorable topology index within
its structural type) has a considerable region of stability as
well.

In summary, we have demonstrated that the complicated phase behavior
of bicontinuous cubic phases in ternary systems can be understood in
terms of the interplay between their universal geometrical properties
and the concentration constraints of a ternary system.  The main
result is that the gyroid structures are favored since they have the
largest values for the topology index $\Gamma$, that is the smallest
porosity among the bicontinuous cubic phases. It should be pointed
out, however, that this work rests on the basic assumption that all
structures form surfaces of constant mean curvature.  Such surfaces
arise as minima of $\int dA$ under a volume constraint, or as minima
of $\int dA (H - c_0)^2$ without volume constraint, but they are {\it
  not} solutions to $\int dA (H - c_0)^2$ under volume constraint ---
except for the special case of vanishing frustration, where the
surface with $H=c_0$ just satisfies the volume constraint.  It is easy
to understand that, in general, surfaces of constant mean curvature do
not minimize the curvature energy with volume constraint, since the
energy can be lowered by keeping the mean curvature on the largest
part of the surface very close to $c_0$, and by concentrating
deviations from $H=c_0$ --- which are enforced by the volume
constraint --- to a small part of the surface (which therefore makes a
small contribution to the curvature integral).  However, since the
regions of stability of most bicontinuous phases do not extend very
far from the lines of vanishing frustration in the phase diagram, the
deviations of the exact solutions from surfaces of constant mean
curvature can be expected to be small for the physically relevant
regions.

In this work, we focused on the case of finite and positive
spontaneous curvature.  Its value can be controled in amphiphilic
systems by changing temperature and disappears at the balanced
temperature. Above the balanced temperature, the spontaneous curvature
is negative and oil and water have to be interchanged in the
structures and phase diagrams discussed. Then the double structures of
type I are replaced by the double structures of type II (the inverse
phases) and the phase boundaries in the Gibbs triangle run
predominantly towards the oil apex.  For surfactant systems, the
balanced temperature is usually well above room temperature
\cite{a:stre94}.  Although the curvature model presented here predicts
inverse bicontinuous phases above the balanced temperature, they are
probably destroyed by thermal fluctuations in this case. For lipid
systems, the balanced temperature is usually similar to the main
transition temperature, and inverse phases are predicted even at room
temperature.  In accordance with experiments on both kinds of systems,
the model presented in this paper predicts that the lamellar phase
dominates at the balanced temperature. However, since it assumes
finite spontaneous curvature and neglects thermal fluctuations, it
cannot describe the bicontinuous microemulsion phase which for
surfactant systems often coexists with the lamellar phase around the
balanced temperature.

We did not consider the effect of thermal fluctuations or long-ranged 
interactions. The special case of vanishing spontaneous curvature has
been investigated in Ref.~\cite{a:Brui92}. It was found that for the
Canham-Helfrich model, the elastic bulk and shear moduli vanish at
$T=0$. Small higher order curvature terms make these moduli finite, 
but thermal long-wavelength fluctuations with large amplitudes should 
remain. An extension of this type of analysis to systems with spontaneous
curvature has not been attempted so far. Since the spontaneous
curvature introduces a new length scale, many of the results of
Ref.~\cite{a:Brui92} should not apply in this case. However, large
fluctuations on long length scales can still be expected.  
The contribution of the fluctuations to the free-energy density should be
of the form $k_BT\ln(\delta/a)/a^3$, where a is the lattice constant, 
which is determined by the concentrations, and $\delta$ is a molecular 
length of the amphiphile \cite{a:port92,a:Brui92,a:mors94}. 
Therefore, the thermal 
contributions to the free energy of different bicontinuous cubic phases 
should be very similar, and the sequences
of stable bicontinuous cubic phases as a function of concentrations
---  as shown in \fig{PhaseDiagramGibbs} and explained in
terms of their geometrical properties in Sec.~\ref{sec:phasebehavior} 
--- should be observed in
experiments. In fact for the system DDAB-water-styrene a sequence of
double structures was reported which seems to correspond to increasing
topology index \cite{a:stro92}.  For the non-cubic phases, thermal
fluctuations correspond mainly to steric interactions between
fluctuating lamellae and cylinders and translational entropy for
spheres. Extending the model by these contributions would in fact
remove the non-concavity of their free energy densities.  Previous
work suggests that the main effect for surfactant systems would be to
favor micellar phases near the water-amphiphile side of the phase
diagram and lamellar phases near the water-apex \cite{a:schw96}.  Such
a modification could bring the calculated phase diagrams of
\fig{PhaseDiagramGibbs} in good qualitative agreement with
experimental phase diagrams measured for the whole Gibbs triangle,
like the ones obtained in Ref.~\cite{a:leav95} for
$H_2O/C_{10}/C_{12}E_5$. For a more detailed comparison, experimental
structure determination for bicontinuous cubic phases in ternary
surfactant systems is needed. The same holds true for ternary lipid
systems where to our knowledge hardly any experimental data is known.

With respect to long-ranged interactions, it seems reasonable to
assume similar effects for all bicontinuous cubic phases.  For
attractive/repulsive forces, we therefore expect all of them to be
favored/disfavored to a similar degree. In principle such additional
contributions could stabilize the bicontinuous cubic phases with
negative curvature index.  In regard to the non-cubic phases, it was
shown previously that van der Waals interactions should show a
considerable effect only for large values of the bending rigidity,
i.e.\ for lipids \cite{a:schw96}.  Thus we predict that the
geometrical arguments presented in this work describe the main physics
of bicontinuous cubic phases in ternary amphiphilic systems
with spontaneous curvature, even when more complicated models are
considered.

\textbf{Acknowledgements:} We thank K.~Gro{\ss}e-Brauckmann for
helpful discussions and J.~Schmalzing and H.~Wagner for drawing our
attention to the connection to integral geometry. USS gratefully
acknowledges support by the Minerva Foundation.

\newpage


\cleardoublepage

\begin{table}
\begin{center}
\begin{tabular}{|l|l|l|l|l|l|l|l|} \hline
     & $\chi$ & $A_0$  & $\Gamma_0$ & $v_0$ & $c$ & $c'$ & $c A_0$ \\ \hline
G    & -8  & 3.0914 & 0.7667 & 0.5   & 0.2191 & 0.1901 & 0.6773 \\ \hline
D    & -16 & 3.8378 & 0.7498 & 0.5   & 0.1411 & 0.1465 & 0.5415 \\ \hline
I-WP & -12 & 3.4641 & 0.7425 & 0.536 & 0.1385 & 0.1592 & 0.4798 \\ \hline
P    & -4  & 2.3451 & 0.7163 & 0.5   & 0.2117 & 0.2188 & 0.4965 \\ \hline  
F-RD & -40 & 4.7707 & 0.6573 & 0.532 & 0.0665 & 0.0906 & 0.3173 \\ \hline  
C(P) & -16 & 3.5105 & 0.6560 & 0.5   & 0.0466 & 0.1226 & 0.1636 \\ \hline  
\end{tabular}
\end{center}
\caption{Properties of the families of triply periodic surfaces of
  constant mean curvature, which are related to their minimal surface
  members.  $\chi$ and $A_0$ are the Euler characteristic and the
  surface area in the conventional unit cell (the value of $A_0$ is
  known exactly in terms of elliptic functions for G, D, P, C(P) and
  I-WP \protect\cite{a:scho70,a:cvij94}).  $\Gamma_0 = ({A_0}^3 / 2 \pi
  |\chi|)^{1/2}$ is the topology index; the structures are ordered
  with with respect to decreasing magnitude of $\Gamma_0$.  $v_0$ is
  the volume fraction of one of the two labyrinths; the volume
  fraction of the other one follows as $1 - v_0$.  $c =
  dv(H^*)/dH^*|_{H^* = 0}$, where $v(H^*)$ is the volume fraction of
  one of the two labyrinths for the corresponding family of surfaces
  of constant mean curvature.  The values for D, I-WP and P are taken
  from Ref.~\protect\cite{a:ande90}, the ones for G, F-RD and C(P) are
  obtained from our spline interpolation of the numerical data of
  Ref.~\protect\cite{a:ande90}.  $c' = - {A_0}^2 / 2 \pi \chi$ is the estimate
  of $c$ as derived in the text.  Note that $c A_0$ gives a very
  similar hierarchy as $c' A_0 = {\Gamma_0}^2$.}
\label{DataTable}
\end{table}

\begin{table}
\begin{center}
\begin{tabular}{|l|l|l|l|} \hline
structure  & labyrinth~1  & surface                      & labyrinth~2 \\ \hline 
Q          & water        & monolayer                    & oil         \\ \hline 
Q${}_{I}$  & oil          & water-filled bilayer         & oil         \\ \hline 
Q${}_{II}$ & water        & oil-filled bilayer           & water       \\ \hline 
\end{tabular}
\end{center}
\caption{Distribution of oil, water and amphiphile in the different
  structural types of bicontinuous cubic phases.  Q denotes single
  structures, Q${}_{I}$ double structures of type I (oil-in-water) and
  Q${}_{II}$ double structures of type II (water-in-oil or inverse).  Although
  double structures consist of two amphiphilic monolayers, for this
  classification we consider each structure to consist of one triply
  periodic minimal surface separating two percolating labyrinths.}
\label{singledouble}
\end{table}

\begin{figure}
\caption{Correspondence between (v,w)-plane and Gibbs triangle.
  Only a certain part of the $(v,w)$-plane is mapped onto the Gibbs
  triangle. The mapping depends on the values for the amphiphile chain
  length $\alpha$ and for the spontaneous curvature $c_0$. Throughout
  this work we use $\alpha = 1/2$ (length is measured in units of
  amphiphile length).  For $c_0 = 1/6$ and $c_0 = 1/12$, the darkly
  and lightly shaded parts are mapped onto the Gibbs triangle,
  respectively. Water, amphiphile and oil apex of the Gibbs triangle
  are denoted by W, A and O, respectively.}
\label{Mapping}
\end{figure}

\begin{figure}
\caption{Minimal surface member  
  for the following families of triply periodic surfaces of constant
  mean curvature: (a) D, (b) C(P), (c) I-WP and (d) F-RD. Shown is
  one conventional unit cell. For single
  structures with vanishing mean curvature, these surfaces represent
  the oil-water interfaces.}
\label{PicturesSingle}
\end{figure}

\begin{figure}
\caption{Geometrical data for triply periodic surfaces of constant
  mean curvature: (a) scaled surface area $A^*$ and (b) scaled mean
  curvature $H^*$ as a function of volume fraction $v$.  Each
  family exists only for a certain $v$-interval; it can be used in (b)
  to identify the family corresponding to each curve. We only show the two
  branches connected by the minimal surface member; there exist other
  branches within the same $v$-intervals which correspond to dense
  arrangements of nearly spherical regions connected by small necks.
  They are assumed to have no physical relevance. The two branches
  used are symmetrical for G, D, P and C(P) since their minimal
  surface members are balanced.}
\label{TPHS}
\end{figure}

\begin{figure}
\caption{(a) Single and (b) double structure
  of the G-family. The two gyroid structures are the most
  stable bicontinuous cubic phases for negative saddle splay modulus
  $\bar \kappa$. For the double gyroid of type I, the water forms the
  sheet-like region between the two monolayers which effectively form
  a bilayer in (b).  For the double gyroid of type II, it fills the
  two channel networks which are separated by this bilayer.}
\label{PicturesDouble}
\end{figure}

\begin{figure}
\caption{(a) Curvature index $\Lambda$ and (b) topology index $\Gamma$
  as a function of volume fraction $v$ for all structures considered.
  Although the different structures can be identified according to
  their $v$-interval of existence (compare Fig.~\ref{TPHS}), here we
  intend to demonstrate only the difference between the three
  structural types.  For Q (solid lines), Q${}_{I}$ (dashed lines) and
  Q${}_{II}$ structures (dotted lines) the curvature index vanishes
  and the topology index attains its maximal value for $v = v_0$, $v =
  1$ and $v = 0$, respectively. Q${}_{II}$ structures and Q structures
  with $v > v_0$ have negative curvature indices and therefore are not
  stable for $c_0 > 0$.}
\label{IndexesPlot}
\end{figure}

\begin{figure}
\caption{Lines of vanishing frustration of various structures for $r = 0$.  
  In the $(v,w)$-plane, the lines $w(v)$ correspond to values for
  $(w,v)$ where a specific phase has vanishing free-energy density.
  For $w \gg 1$ the Q and Q${}_{I}$ structures are stable
  for $v \lesssim v_0$ and $v \lesssim 1$,
  respectively. The lines at $w = 3$ and $w = 4$ correspond to spheres
  and cylinders, respectively. For $c_0 = 1/6$, only the part below
  the dotted line is mapped onto the Gibbs triangle.}
\label{PhaseDiagramRfixed}
\end{figure}

\begin{figure}
\caption{Phase diagram as a function of $w$ and $r$ for (a) $v =
  0.1$, (b) $v = 0.2$, (c) $v = 0.4$ and (d) $v = 0.6$. For $c_0 =
  1/6$, only the region below the dashed line is mapped onto the Gibbs
  triangle. The most stable phase with respect to $r$ is the
  double gyroid G${}_{I}$ which however cannot exist for $v < 0.112$.}
\label{PhaseDiagramVfixed}
\end{figure}

\begin{figure}
\caption{Phase behavior in the Gibbs triangle for (a) $r
  = 0.01$, $c_0 = 1/6$, (b) $r = 1/15 = 0.067$, $c_0 = 1/6$ and (c) $r
  = 1/15$, $c_0 = 1/12$. For larger values of $r$, the gyroid
  structures G and G${}_{I}$ dominate. Lowering spontaneous curvature
  $c_0$ corresponds to raising temperature and extends the region of
  stability for the lamellar phase L. For simplicity, in these plots
  we do not consider the close-packing constraint for S.}
\label{PhaseDiagramGibbs}
\end{figure}

\psfig{file=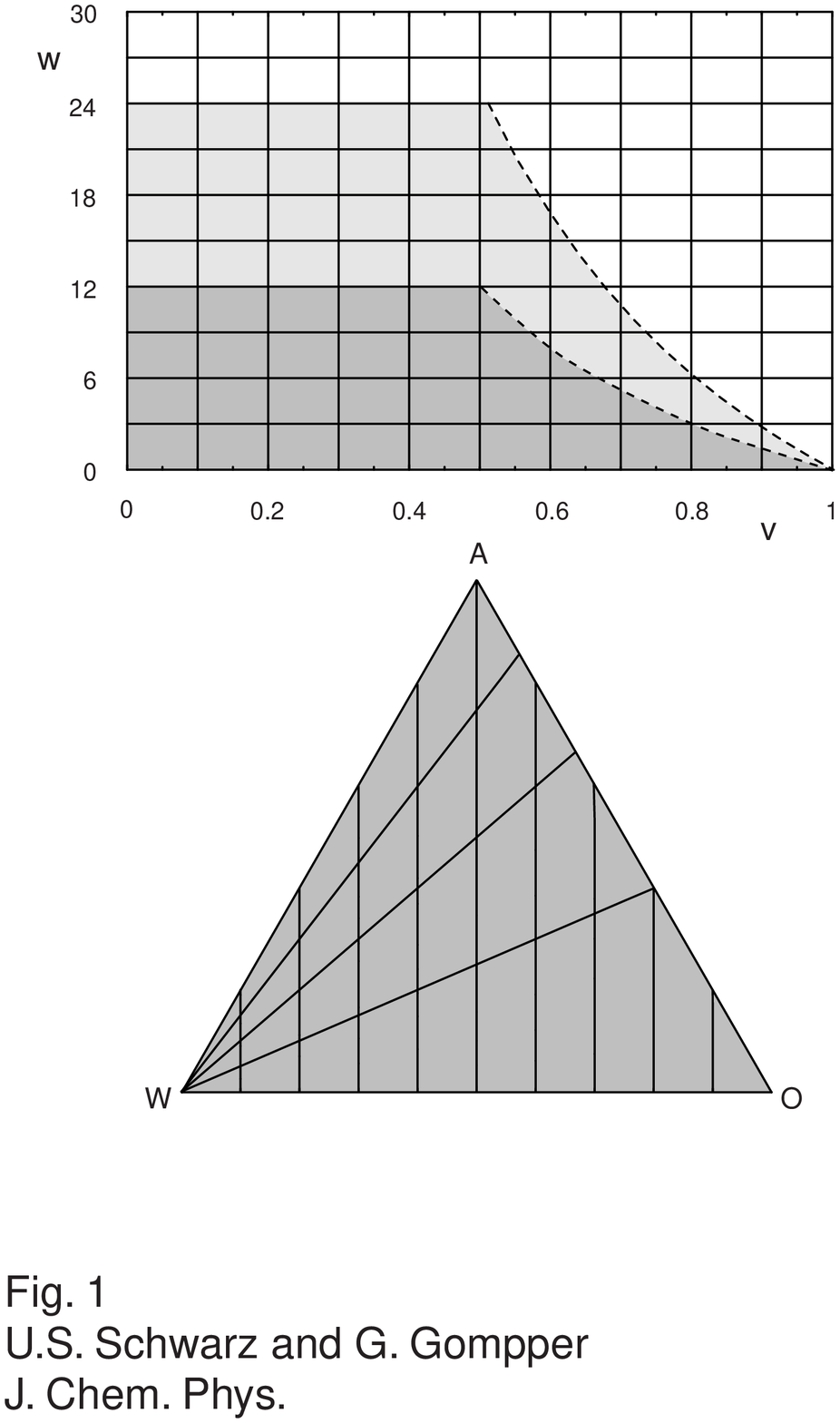}
 
\psfig{file=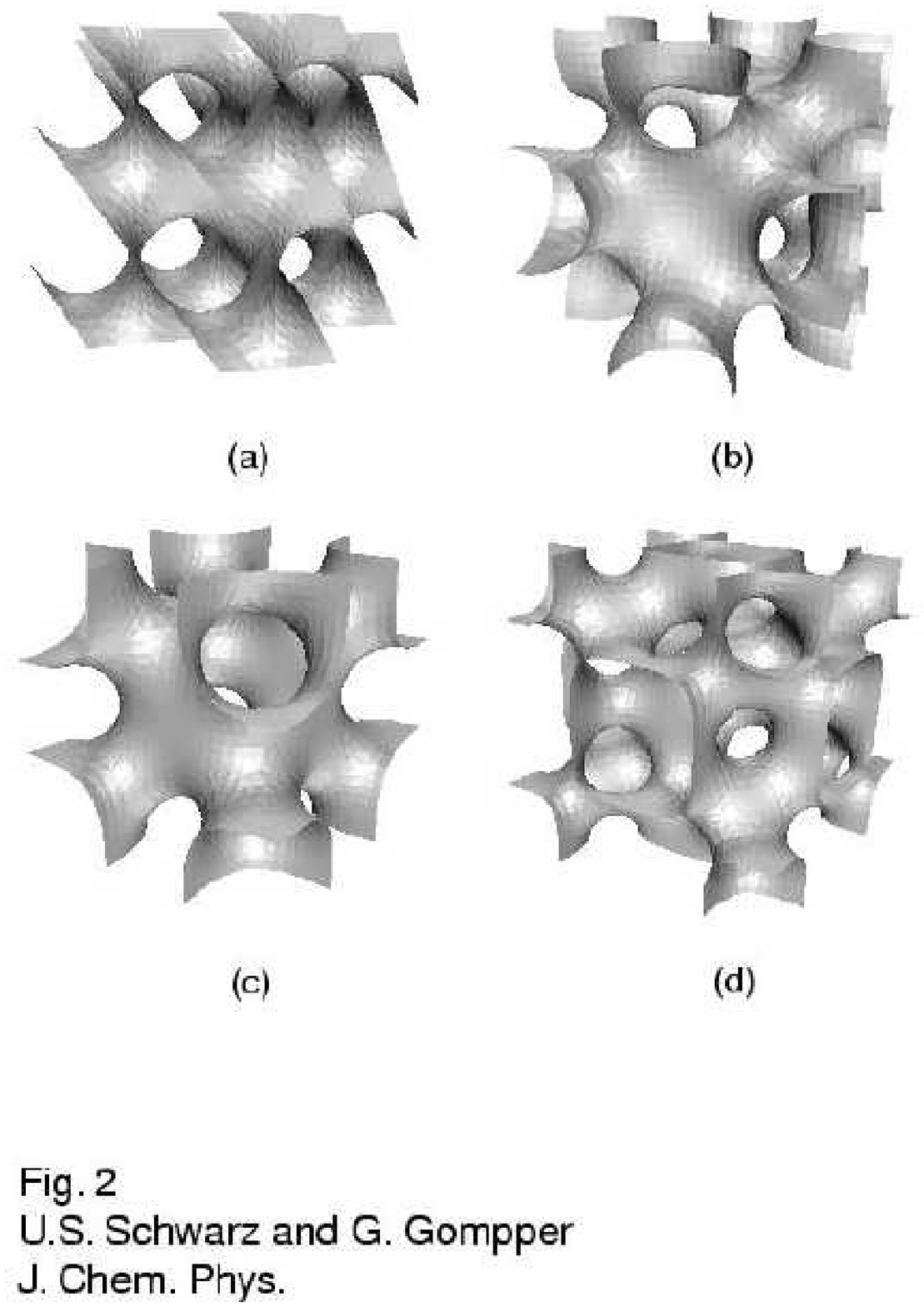}

\psfig{file=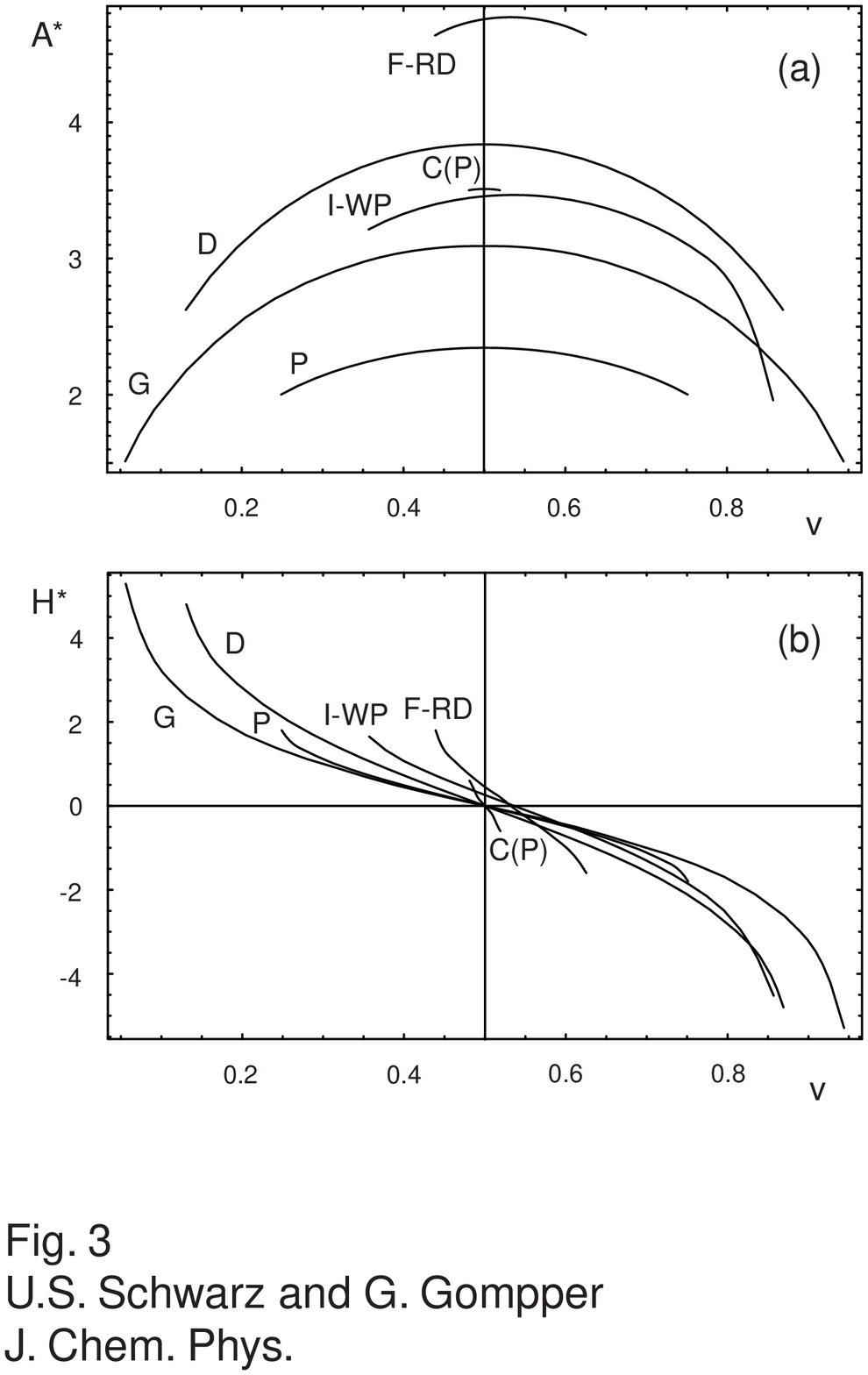}

\psfig{file=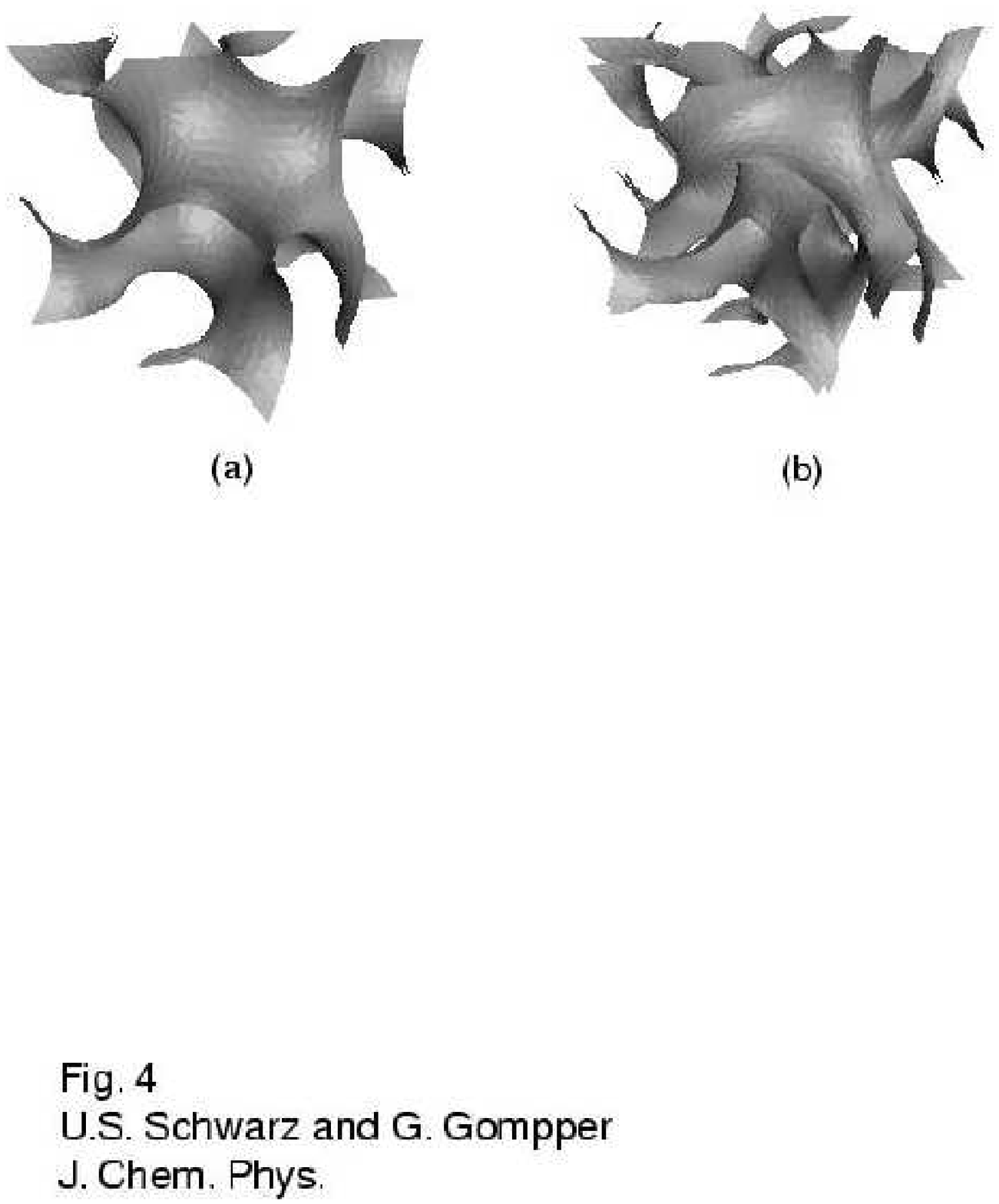}

\psfig{file=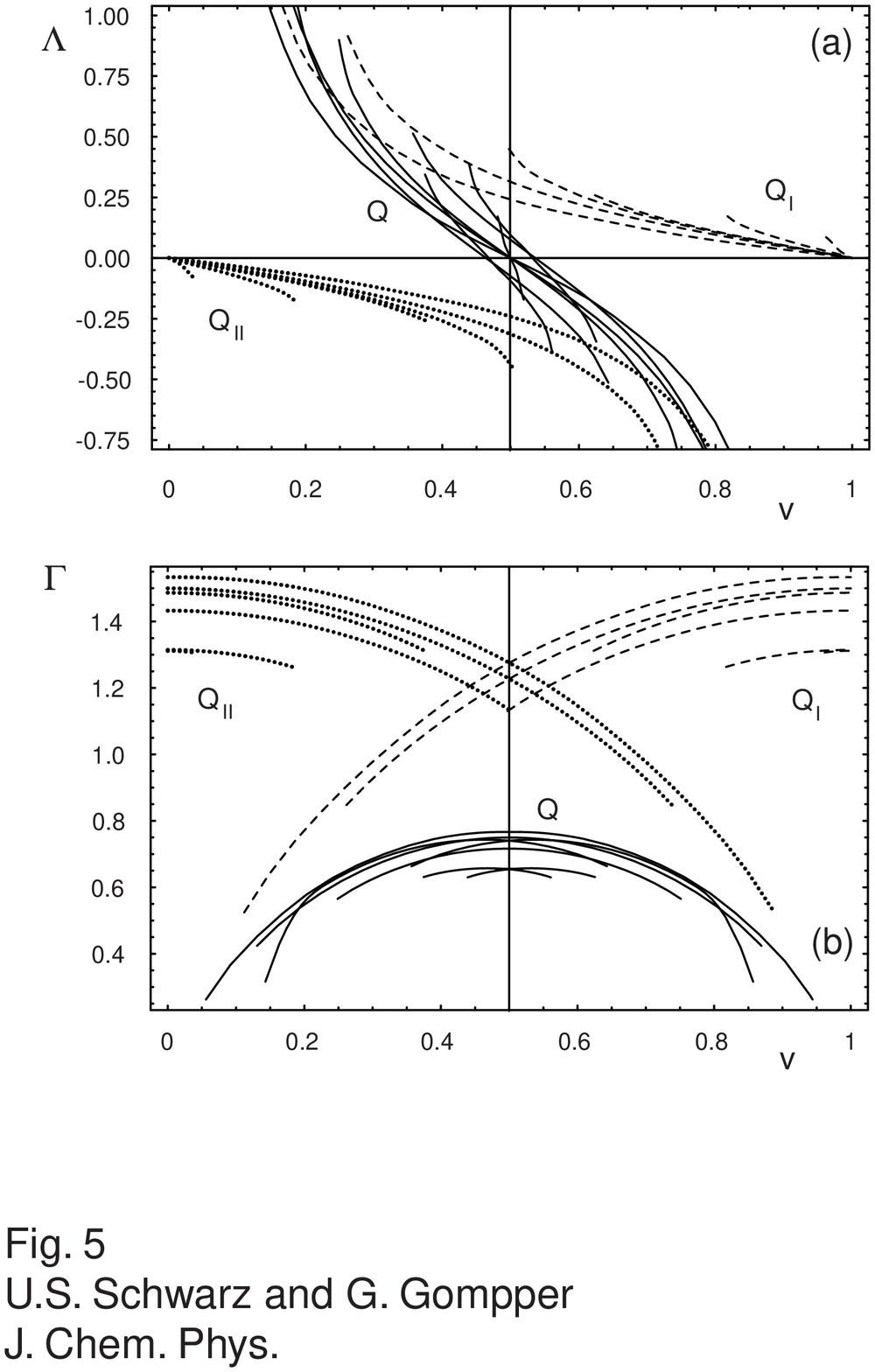}

\psfig{file=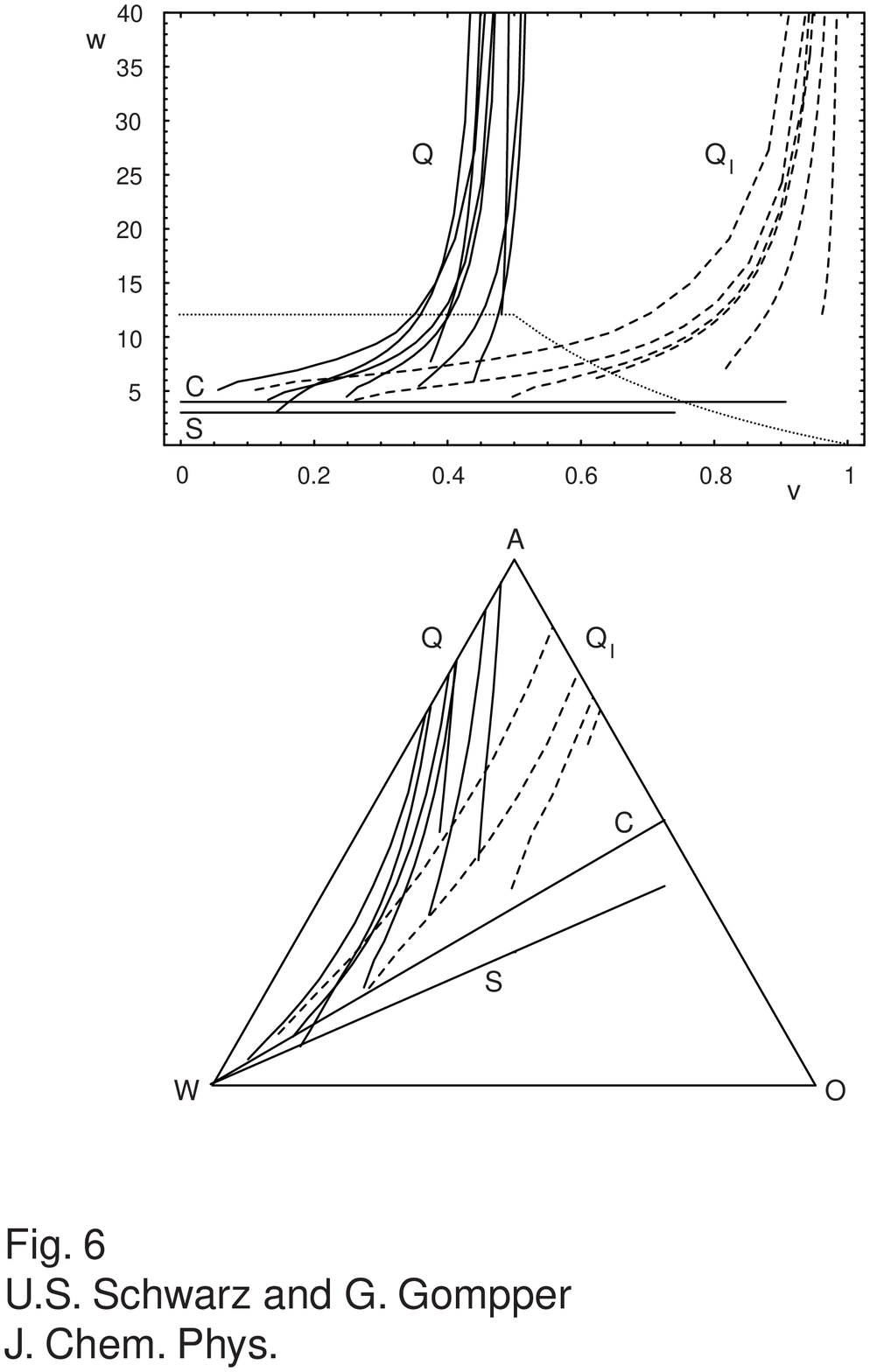}

\psfig{file=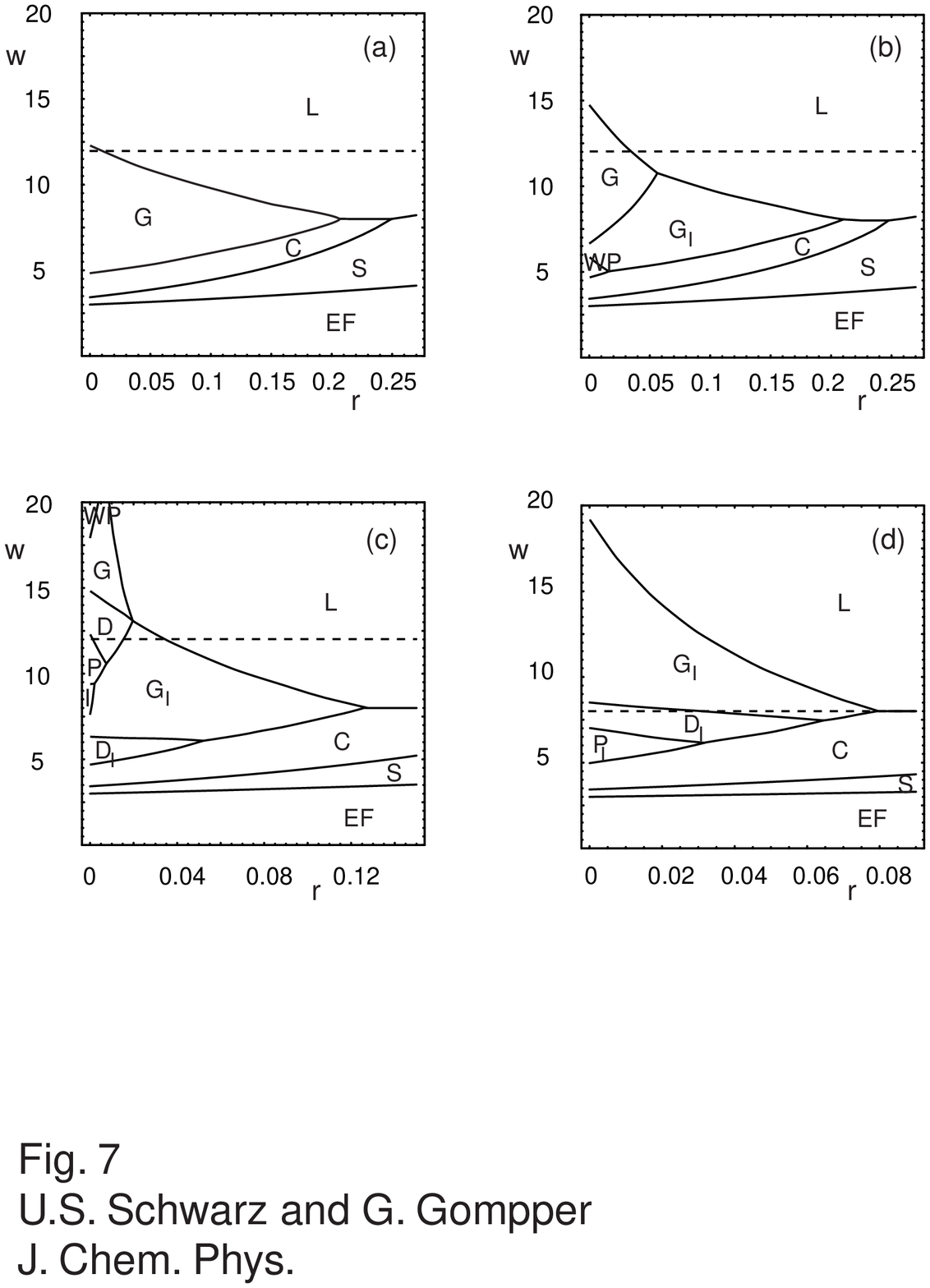}

\psfig{file=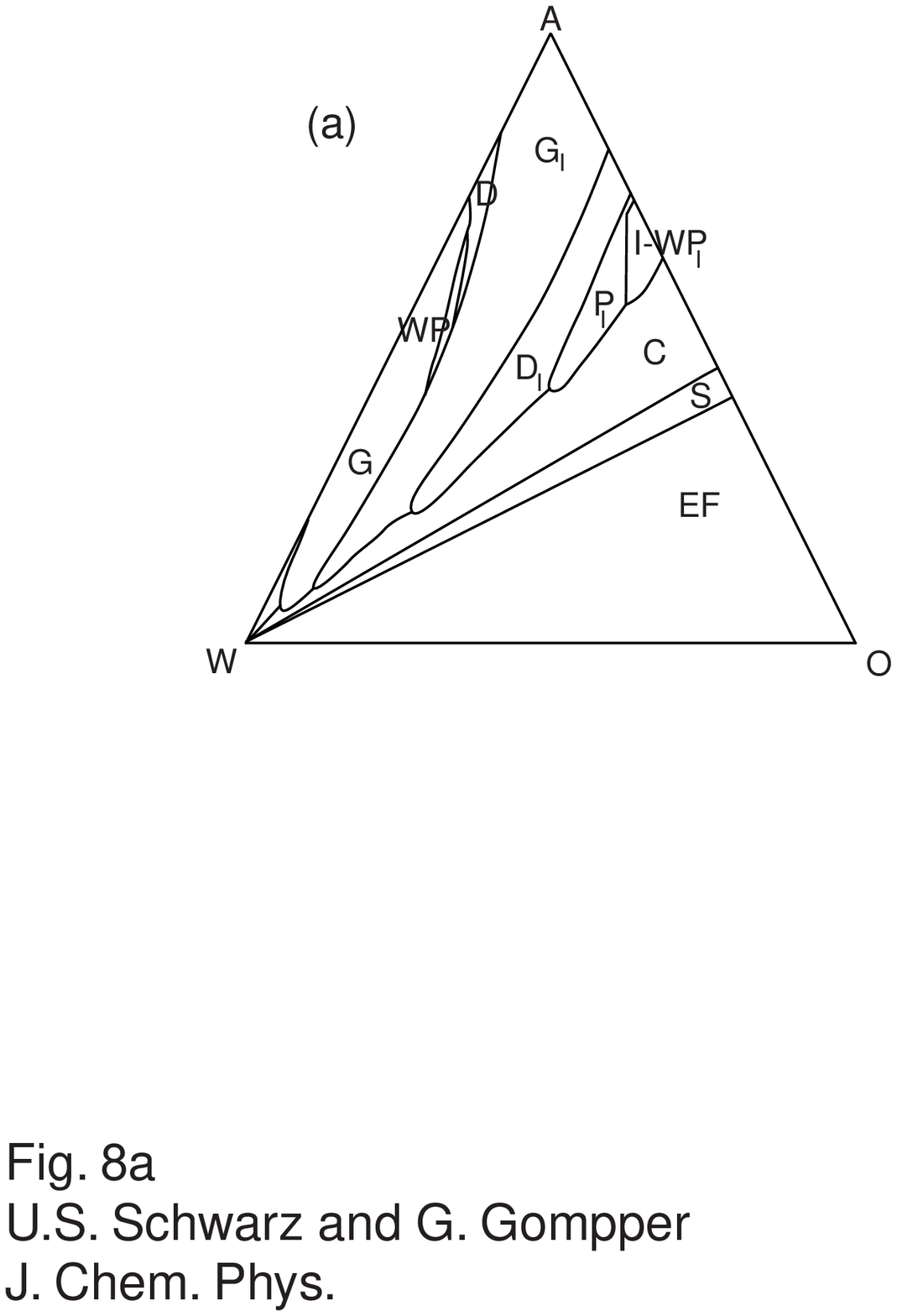}

\psfig{file=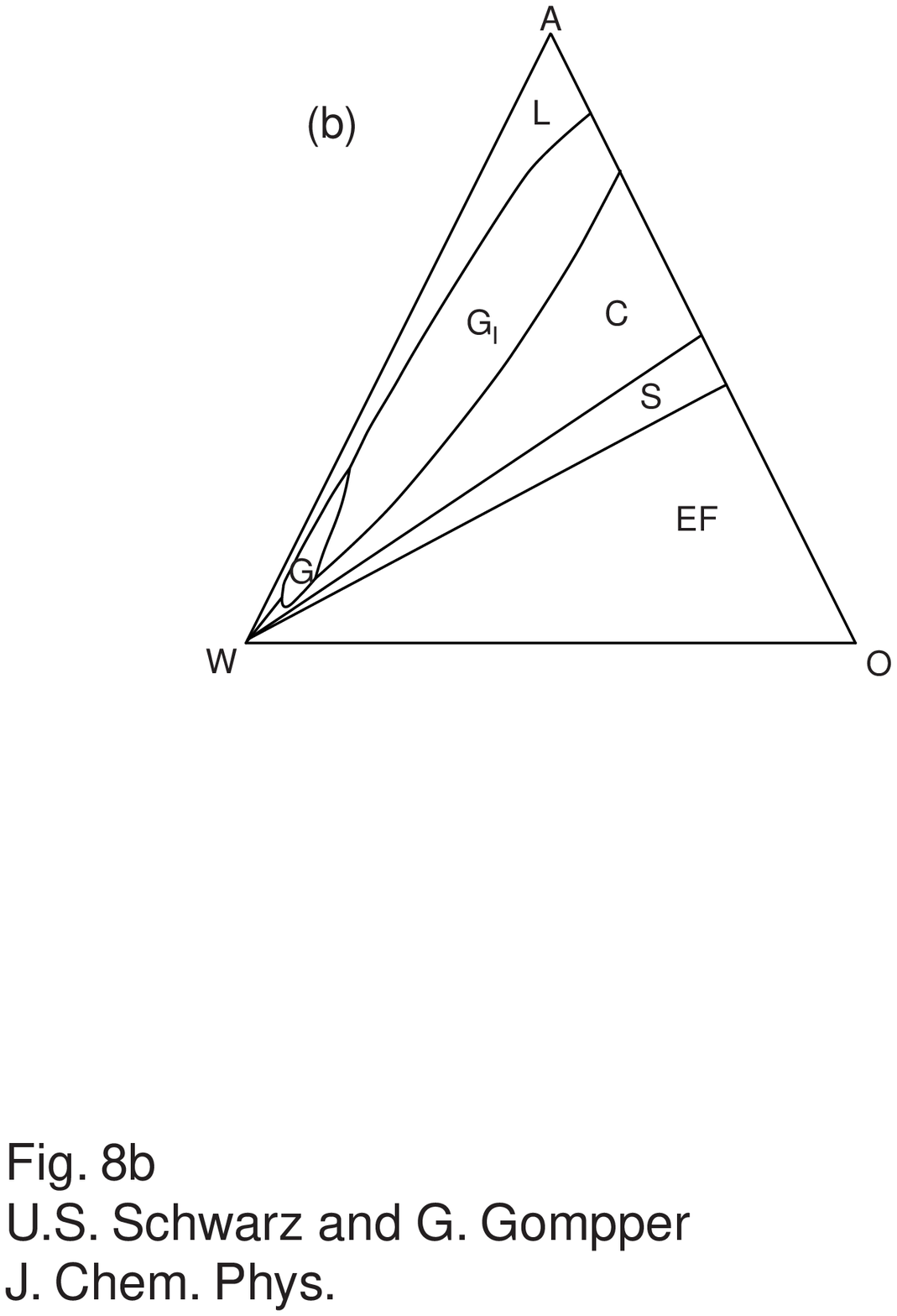}

\psfig{file=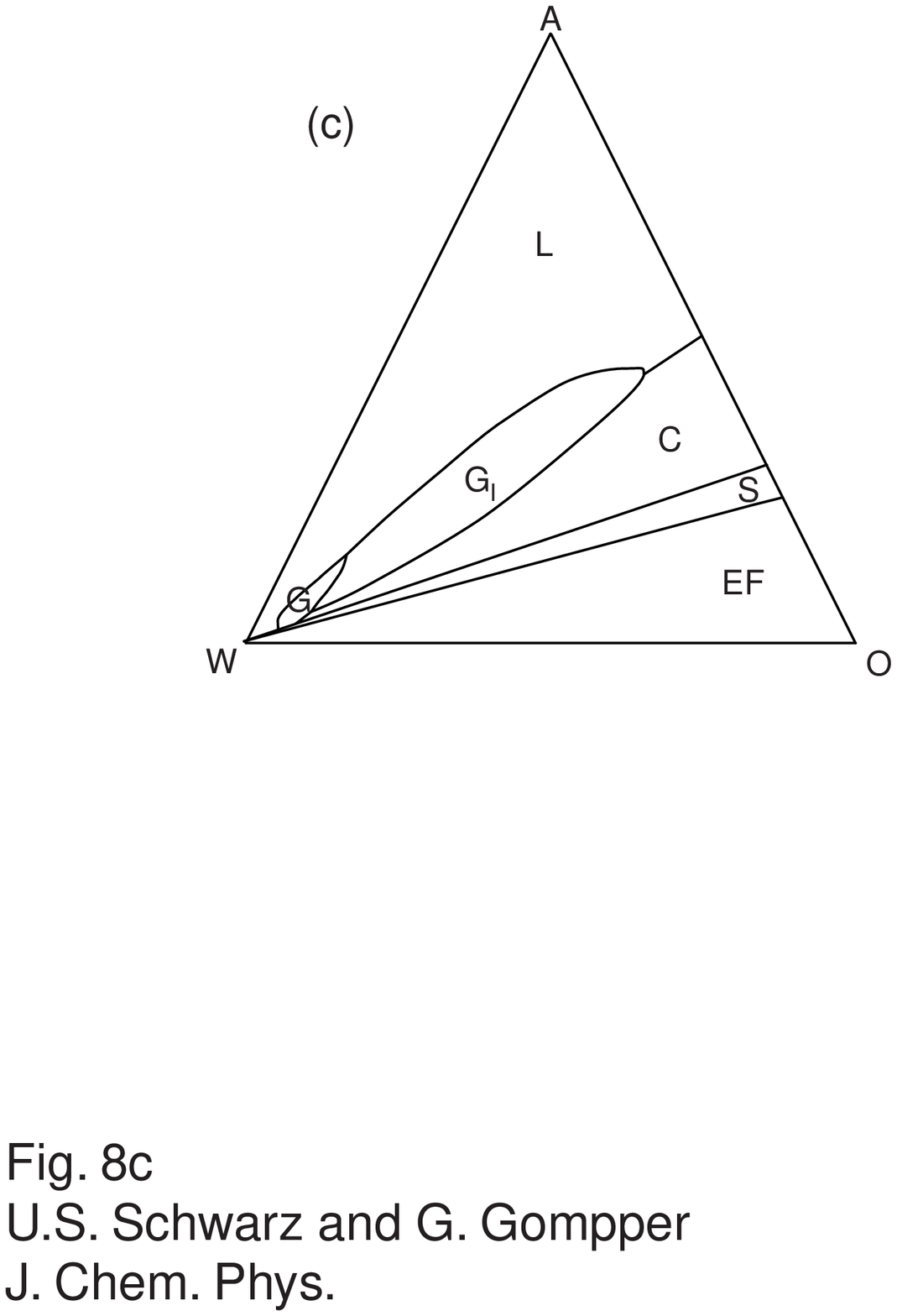}

\end{document}